\documentclass[12pt]{article}
\usepackage[utf8]{inputenc}
\usepackage{amsmath,amssymb,aas_macros}
\usepackage[dvipdfmx]{graphicx}
\usepackage{color}
\usepackage{cite}
\usepackage{mathrsfs}
\usepackage{bm}
\usepackage{enumerate}
\usepackage{url}

\usepackage[x11names]{xcolor}
\usepackage[whole]{bxcjkjatype}

\usepackage[colorlinks,citecolor=blue]{hyperref}
\renewcommand{\thefootnote}{\#\arabic{footnote}}

\newcommand{\mr}[1]{\mathrm{#1}}	%

\makeatletter
\@addtoreset{equation}{section}
\makeatother

\begin{document}

\begin{titlepage}

\begin{center}

\vskip .75in

{\Large \bf
Impact of the primordial fluctuation  \vspace{1mm} \\
power spectrum on the reionization history
}

\vskip .75in

{\large
Teppei Minoda$\,^{a}$, Shintaro Yoshiura$\,^{b}$, Tomo Takahashi$\,^c$
}

\vskip 0.25in

{\em
$^{a}$Tsinghua University, Department of Astronomy, Beĳing 100084, China 
\vspace{2mm} \\
$^{b}$Mizusawa VLBI Observatory, National Astronomical Observatory Japan, 2-21-1 Osawa, Mitaka, Tokyo 181-8588, Japan
\vspace{2mm} \\
$^{c}$Department of Physics, Saga University, Saga 840-8502, Japan 
}

\end{center}
\vskip .5in

\begin{abstract}
We argue that observations of the reionization history can be used as a probe of primordial density fluctuations, particularly on small scales. 
Although the primordial curvature perturbations are well constrained from measurements of cosmic microwave background (CMB) anisotropies and large-scale structure, these observational data probe the curvature perturbations only on large scales, and hence its information on smaller scales will give us further insight on primordial fluctuations. Since the formation of early galaxies is sensitive to the amplitude of small-scale perturbations, and then, in turn, gives an impact on the reionization history, one can probe the primordial power spectrum on small scales through observations of reionization.
In this work, we focus on the running spectral indices of the primordial power spectrum to characterize the small-scale perturbations, and investigate their impact on the reionization history using the numerical code \texttt{21cmFAST}, which adopts a simple but commonly used reionization model.
We also derive the constraints on the running spectral indices from observations of the reionization history indicated by the luminosity function of the Lyman-$\alpha$ emitters.
We show that the reionization history, in combination with large-scale observations such as CMB, would be a useful tool to investigate primordial density fluctuations.
\end{abstract}

\end{titlepage}

\renewcommand{\thepage}{\arabic{page}}
\setcounter{page}{1}
\renewcommand{\thefootnote}{\#\arabic{footnote}}
\setcounter{footnote}{0}

%%%%%%%%%%%%%%%%%%%%%%%%%%%%%%%%%%%%%%%%
\section{Introduction \label{sec:intro}}
%%%%%%%%%%%%%%%%%%%%%%%%%%%%%%%%%%%%%%%%
The statistical property of primordial perturbations is well understood by measurements of cosmic microwave background (CMB) \cite{2020A&A...641A..10P} and large-scale structure (LSS) \cite{2019PASJ...71...43H, 2021A&A...646A.140H, 2022PhRvD.105b3520A, 2020JCAP...04..038P, 2021PhRvD.103h3533A, 2022ApJS..259...35A} in this decade. 
The perturbations are almost adiabatic and Gaussian.
The power spectrum of the primordial perturbations is nearly scale-invariant, but slightly red-tilted. These characteristics of the primordial perturbations provide strong constraints on inflation and their generation mechanism.
%TT \cite{2022PhRvD.105d3504E,2022PhRvD.106f1301E,2023MNRAS.520.2405B}. 

The scale dependence of primordial power spectrum ${\cal P}_\zeta$ is usually characterized by a constant spectral index $n_s$ with which the power spectrum is given by 
\begin{equation}
    {\cal P}_\zeta \propto k^{n_s-1} \,.
\end{equation}
However, in usual inflation models, $n_s$ is not exactly constant, but has some scale dependence, and in some cases, it can be strongly scale-dependent. Such a scale dependence is commonly characterized by the so-called ``runnings" in which the power spectrum is represented by 
\begin{equation}
{\cal P}_\zeta \propto k^{n_s -1 + \frac12 \alpha_s \ln (k /k_0)+ 
\frac16 \beta_s [\ln (k/k_0) ]^2} \,,   
\end{equation}
with $\alpha_s$ and $\beta_s$ being denoted as running parameters at the pivot scale $k_0$.
Because observations such as CMB and LSS probe the primordial fluctuations only on large scales as $10^{-3}~\mr{Mpc}^{-1} \lesssim k \lesssim 1~\mr{Mpc}^{-1}$ \cite{2020A&A...641A..10P}, we need to investigate the fluctuations on different scales in order to study the scale dependence of the primordial fluctuations further. 
Indeed, there have been many attempts to investigate fluctuations on small scales $k \gtrsim 1~\mr{Mpc}^{-1}$ to probe primordial fluctuations using observations such as the 21-cm global signal \cite{2018PhRvD..98f3529Y,2020PhRvD.101h3520Y}, 21-cm power spectrum \cite{2008PhRvD..78b3529M,2013JCAP...10..065K}, 21-cm forest \cite{2014PhRvD..90h3003S}, Lyman-$\alpha$ forest \cite{2011MNRAS.413.1717B,2015JCAP...11..011P}, 21-cm signal from minihalos \cite{2018JCAP...02..053S}, ultracompact minihalos \cite{2012PhRvD..85l5027B,2018PhRvD..97b3539N}, primordial black holes formation \cite{2009PhRvD..79j3520J,2019PhRvD.100f3521S}, supernovae lensing \cite{2016MNRAS.455..552B}, spectral distortions of CMB \cite{1994ApJ...430L...5H,2012MNRAS.425.1129C,2012ApJ...758...76C,2013JCAP...06..026K,2014JCAP...10..046C,2016PhRvD..94b3523C,2017JCAP...11..002K}, galaxy UV luminosity function \cite{2020PhRvD.102h3515Y}, the substructure of dark matter halo \cite{2022PhRvD.106j3014A} and so on.

Our approach in this paper is to study the impact of the primordial power spectrum on the  reionization history. 
Since fluctuations on small scales $k \gtrsim 1~\mr{Mpc}^{-1}$ are connected to the halos with mass $M \lesssim 10^{11} M_\odot$, which are expected to be formed around the cosmic dawn and the epoch of reionization, and thereby we can utilize observations of the reionization history as a probe of such small-scale primordial fluctuations.
One famous cosmological constraint on the reionization history is the Thomson optical depth of CMB, which corresponds to the column density of the free electron. Therefore, it provides the redshift-integrated information on the reionization history. 
We can also probe the reionization physics from observations of the Lyman-$\alpha$ emitter (LAE) luminosity function 
which gives the redshift-dependent information on the reionization history. We focus on the latter observation to probe the primordial fluctuations from the reionization history by using a simple and widely-used numerical calculation code \texttt{21cmFAST}. In particular, we use reionization history estimated by the SILVERRUSH survey data with the Subaru telescope \cite{2021ApJ...923..229G}, from which we can constrain the running of the spectral index $\alpha_s$ and $\beta_s$.

In the following, we first introduce observational constraints on the reionization history in Section~\ref{sec:2-1}, and our method to calculate the evolution of the free electron fraction in Section~\ref{sec:2-2}.
Then in Section~\ref{sec:2-3}, we provide the constraints on the runnings of the primordial power spectrum from observations of the reionization. We conclude and summarize this paper in Section~\ref{sec:3}. 
In our analysis, we fix the cosmological parameters, other than the runnings of the spectral index, as $\sigma_8=0.831, H_0=67.27$ km/s/Mpc, $\Omega_\mr{m}=0.32, \Omega_\mr{b}=0.049,$ and $n_s=0.9586$, assuming the flat $\Lambda$CDM cosmology \cite{2016A&A...594A..13P}.

\section{Reionization history as a probe of primordial fluctuations}

\subsection{Observational constraint on the reionization history}
\label{sec:2-1}
In this work, we put a constraint on the running parameters of the primordial power spectrum from observations of the Ly-$\alpha$ luminosity function with SILVERRUSH by calculating the reionization history \cite{2021ApJ...923..229G}. 
Indeed one can also use the measurement of the CMB optical depth from Planck \cite{2020A&A...641A...6P} to obtain the constraint. However, as we show in the following, the constraint on the optical depth from the Ly-$\alpha$ luminosity function with SILVERRUSH is tighter than that from CMB, and hence we focus on the constraint from the reionization history (more specifically the value of $x_e$ at some redshift) estimated by the Ly-$\alpha$ luminosity function.
Firstly, by using a simple $\tanh$-type reionization model, we demonstrate how the reionization history is constrained by the latest Planck and SILVERRUSH data. The free electron fraction at a given redshift $z$ is often modeled by the hyperbolic tangent function as \cite{2008PhRvD..78b3002L}
\begin{align}
x_\mr{e}(z, z_\mr{re})=\cfrac{1}{2} \left[1+\tanh{\left(\cfrac{y(z_\mr{re})-y(z)}{\Delta y}\right)}\right],
\label{eq:tanh_reionization}
\end{align}
where $z_\mr{re}$ is the model parameter that represents the central redshift of reionization, and $y(z)=(1+z)^{2/3}, ~\Delta y= 3/2 (1+z_\mr{re})^{1/2}\Delta z$, with $\Delta z=0.5$.
We note that although the redshift width of reionization $\Delta z$ is in general considered to be a free parameter as assumed in the original reference \cite{2008PhRvD..78b3002L}, we fix it as $\Delta z=0.5$ in most of our analyses since this width fits the SILVERRUSH results well and the recent Planck analysis also assumes $\Delta z=0.5$ \cite{2020A&A...641A...6P}. 
Only in Section~\ref{sec:2-4}, we vary $\Delta z$ to argue that the width $\Delta z$ does not actually affect the constraints very much.

\begin{figure}
    \centering
	\includegraphics[width=0.8\columnwidth]{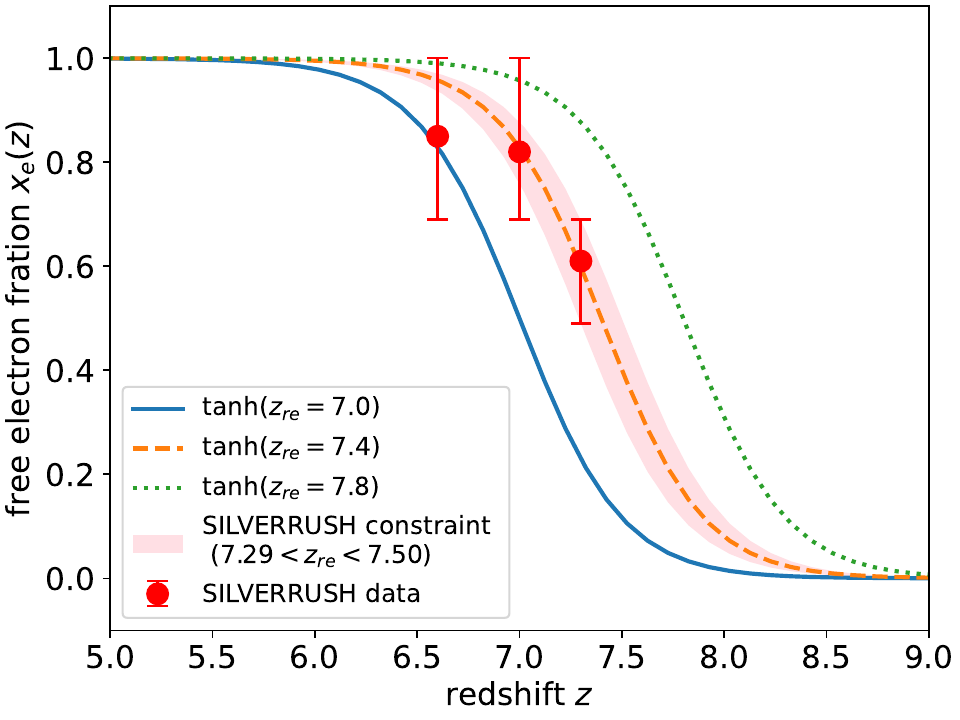}
    \caption{
    Redshift evolution of the free electron fraction with $z_\mr{re}=7.0$ (solid), $7.4$ (dashed), and $7.8$ (dotted). Three red points at $z=6.6, 7.0,$ and $7.3$ are the SILVERRUSH results, given in Table 2 of Ref. \cite{2021ApJ...923..229G}. The pink-shaded region corresponds to the range of $x_\mr{e}$ from the constraint of the SILVERRUSH results for the $\tanh$-type reionization model with $7.29 \le z_\mr{re} \le 7.50$.
    }
    \label{fig:IonHist_calc_SR}
\end{figure}

Figure \ref{fig:IonHist_calc_SR} shows the $\tanh$-type reionization history for different $z_\mr{re}$, together with the constraint on the free electron fraction from the SILVERRUSH results of the Ly-$\alpha$ luminosity function. 
Three data points from the SILVERRUSH results are depicted in the figure. The most stringent constraint among them is $0.49 \le x_\mr{e} \le 0.69$ at $z=7.3$. This constraint can be translated into that on the central redshift of the reionization $z_\mr{re}$ as $7.29 \le z_\mr{re} \le 7.50$ for the modeling of $x_\mr{e}$ given in Eq.~\eqref{eq:tanh_reionization}. 
The corresponding range for $x_\mr{e}$ is shown with the pink-shaded region in Figure \ref{fig:IonHist_calc_SR}. We note that, even when we change the redshift width of the reionization as $\Delta z=1.0$, the constraint on the central redshift of reionization is given by $7.42 \le z_\mr{re} \le 7.70$, which is consistent with that for the case of $\Delta z=0.5$. Therefore the change of $\Delta z$ does not give a significant impact on our conclusion.
We show the case of varied $\Delta z$ in section \ref{sec:2-4}.

Other SILVERRUSH data points, $x_\mr{e} \ge 0.69$ at $z=7.0$ and $x_\mr{e} \ge 0.69$ at $z=6.6$, can be translated into $z_\mr{re} \ge 7.20$ and $z_\mr{re} \ge 6.80$, respectively. These constraints are weaker than that from the data point at $z=7.3$, so we do not use two data points at $z=7.0$ and 6.6 in the following analysis unless explicitly mentioned.

\begin{figure}
    \centering
	\includegraphics[width=0.7\textwidth]{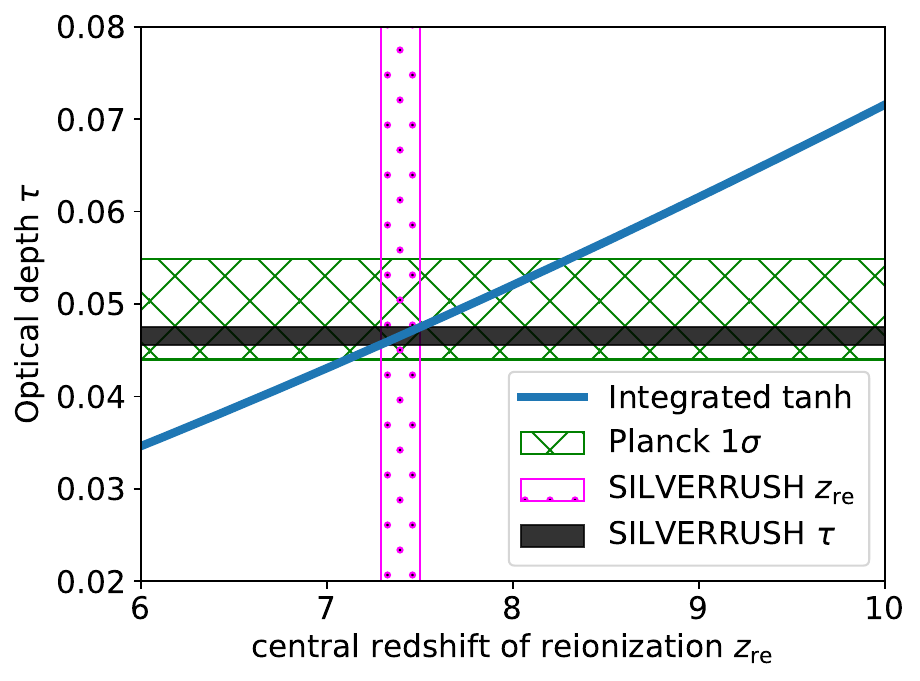}
    \caption{Thomson optical depth obtained by Eq. \eqref{eq:optical_depth} with the $\tanh$-type reionization history, as a function of $z_\mr{re}$ (the blue thick line). The green cross-hatched horizontal region is the constraint on the optical depth from Planck 2018 results. The magenta dot-hatched vertical one is obtained by the $\tanh$-type reionization history with the SILVERRUSH results, as $7.29 \le z_\mr{re} \le 7.50$. The horizontal thick-shaded region corresponds to the range of $\tau$ suggested from the SILVERRUSH constraints on $z_\mr{re}$ for the $\tanh$-type reionization model.}
    \label{fig:OpticalDepth_calc_Planck}
\end{figure}

To compare the constraint from SILVERRUSH with that from the Planck optical depth, we calculate the Thomson optical depth, which is obtained by integrating the $\tanh$-type reionization history, via
\begin{align}
\tau=n_{\mathrm{H}} c \sigma_{\mathrm{T}} \int_0^{z_{\max}} dz~x_{\mathrm{e}}(z, z_\mr{re}) \frac{(1+z)^2}{H(z)},
\label{eq:optical_depth}
\end{align}
with the number density of hydrogen $n_{\mathrm{H}}$, the speed of light $c$, the Thomson scattering cross-section $\sigma_{\mathrm{T}}$, and the Hubble parameter $H(z)$. As referred in \cite{2020A&A...641A...6P}, we take the maximum redshift in the integration as $z_{\max}=50$. We show the Thomson optical depth from the $\tanh$-type reionization with different $z_\mr{re}$ in Figure \ref{fig:OpticalDepth_calc_Planck}.
The SILVERRUSH constraint on the central redshift of the reionization is $7.29 \le z_\mr{re} \le 7.50$, and plotted with the dot-hatched region in Figure \ref{fig:OpticalDepth_calc_Planck}. Assuming the $\tanh$-type reionization history, this constraint on $z_\mr{re}$ can be translated to the constraint on the Thomson optical depth, as $0.0456 \le \tau \le 0.0475$, which is obtained from the overlapping range between the blue thick line and the magenta dot-hatched region. This constraint on $\tau$ is plotted with the horizontal thick-shaded region. 
For comparison, we also show the Planck constraint on the optical depth, $0.0440 \le \tau \le 0.0549$ given in Eq.~(86a) in Ref. \cite{2020A&A...641A...6P} with the horizontal green cross-hatched region. It is clear that the SILVERRUSH constraint is stronger than the Planck one when we assume the $\tanh$-type reionization.

\subsection{Reionization calculation with the \texttt{21cmFAST} \label{sec:2-2}}
\begin{figure}
    \centering
	\includegraphics[width=0.6\textwidth]{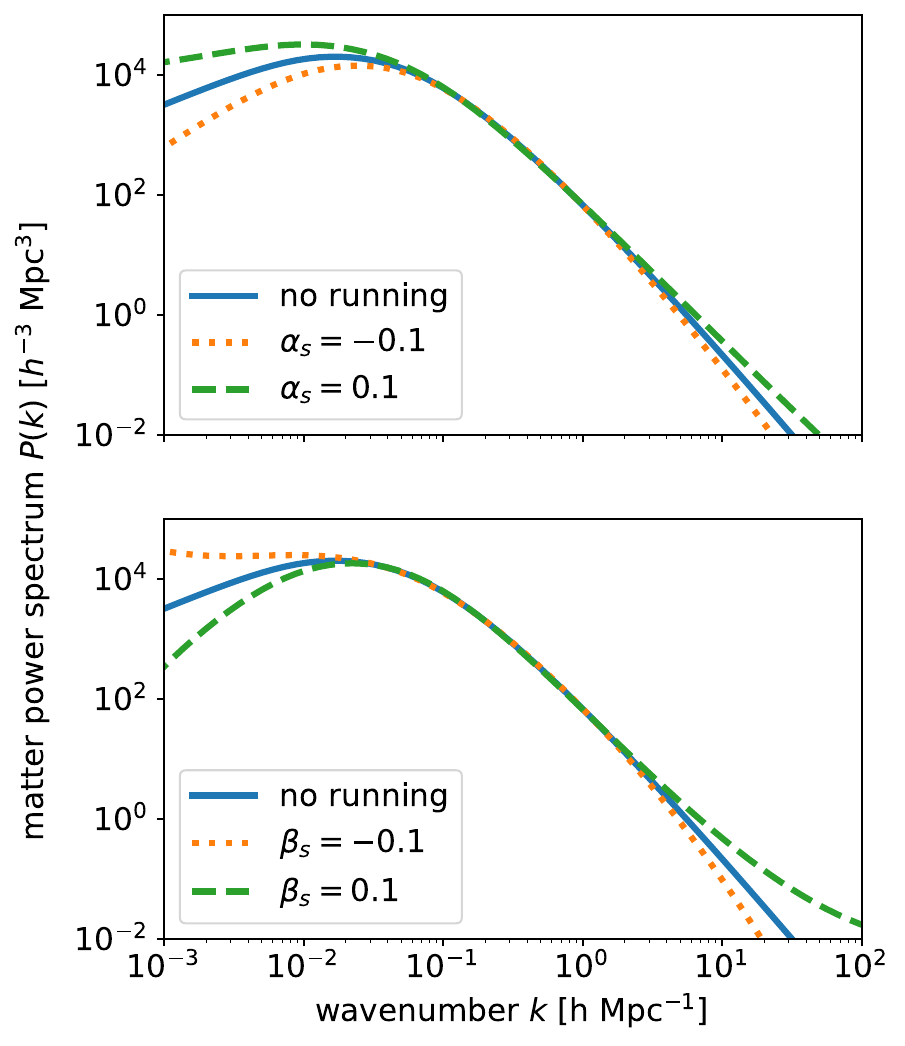}
    \caption{Effects of the running spectral indices $\alpha_s$ and $\beta_s$ on the matter power spectra.}
    \label{fig:PSmat_running_ab}
\end{figure}

\begin{figure}
    \centering
	\includegraphics[width=0.6\textwidth]{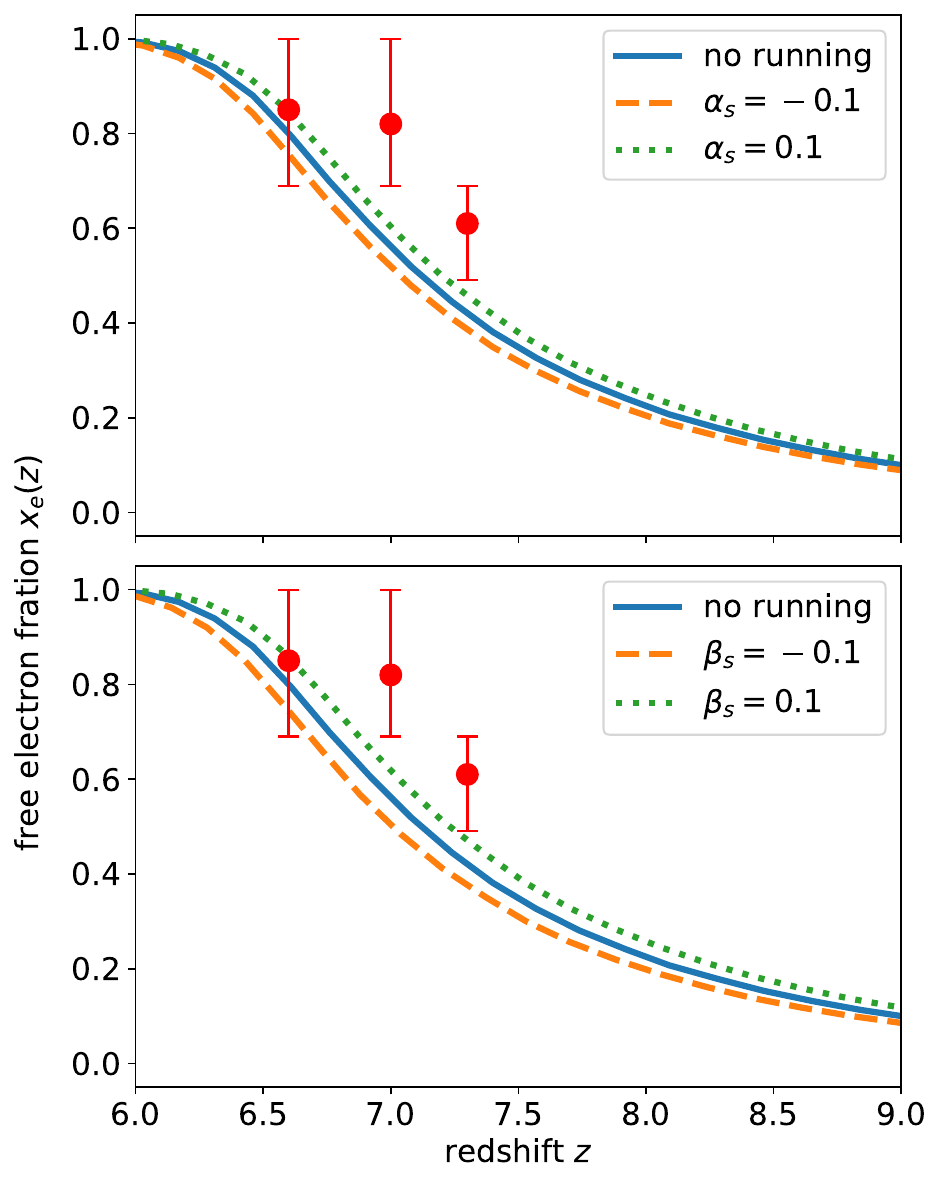}
    \caption{Effects of the running spectral indices $\alpha_s$ and $\beta_s$ on the reionization history. Astrophysical parameters are assumed to be the best-fitted values from HERA constraints~\cite{2022ApJ...924...51A}. 
    Although the HERA best-fitted model slightly deviates from the SILVERRUSH constraints, taking account of the uncertainties, the HERA results are consistent with the SILVERRUSH (see Fig.~\ref{fig:IonHist_3astros}).}
    \label{fig:IonHist_running}
\end{figure}

We adopt the publicly available semi-numerical simulation code \texttt{21cmFAST} \cite{2019MNRAS.484..933P} to calculate the reionization history.
We modify the code to extend the cosmological model in \texttt{21cmFAST}, to include the ``running'' and ``running of running'' of the primordial curvature power spectrum, $\alpha_s$ and $\beta_s$ as
\begin{equation}
    \mathcal{P}_\zeta(k) = A_s \left(\cfrac{k}{k_0}\right)
    ^{n_s-1+\frac{1}{2}\alpha_s \ln \left(\frac{k}{k_0}\right)
    +\frac{1}{6}\beta_s \left[\ln \left(\frac{k}{k_0}\right)\right]^2},
	\label{eq:power}
\end{equation}
where $A_s$ is the amplitude at the pivot scale $k_0$, and $n_s$ is the spectral index. 
We fix the pivot scale as $k_0=0.05~\mr{Mpc}^{-1}$ as the Planck collaboration adopted \cite{2020A&A...641A...6P}. To simplify the following analysis, we fix the $\Lambda$CDM parameters with the Planck 2018 best-fitted values \cite{2020A&A...641A...6P} although we vary several astrophysical parameters as we describe below since the reionization history is more affected by astrophysical processes.
Figure \ref{fig:PSmat_running_ab} shows the matter power spectra with varying running parameters. Here, the larger $\alpha_s$ and $\beta_s$ enhance the small-scale power spectrum. However, on large scales, the larger $\alpha_s$ enhances the power spectrum as well, but the larger $\beta_s$ reduces the large-scale matter power spectrum. 
Although cosmological observations such as CMB and LSS can probe large-scale fluctuations, constraints from such observations show the degeneracy between $\alpha_s$ and $\beta_s$. 
However, the combined analysis with other observations especially probing the small-scale fluctuations is expected to break such a degeneracy.
In this paper, we focus on the reionization history as such a cosmological probe for the running spectral indices. Figure \ref{fig:IonHist_running} shows the time evolution of the free electron fraction with different running parameters. 
Increasing $\alpha_s$ and $\beta_s$ enhances the amplitude of small-scale matter fluctuations, and thereby high-redshift halo formation is promoted and the reionization gets earlier.

\begin{figure}
    \centering
	\includegraphics[width=0.7\textwidth]{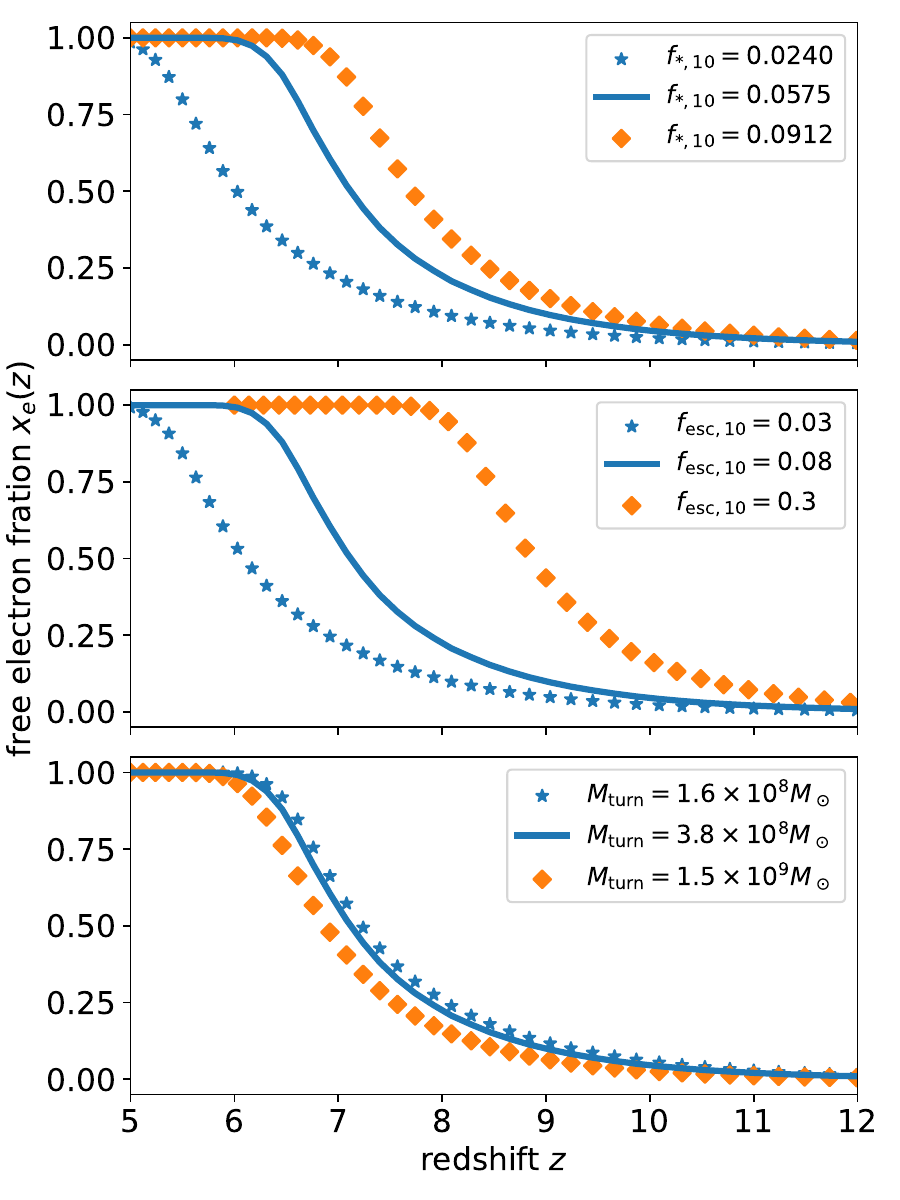}
    \caption{Reionization history calculated by \texttt{21cmFAST} with varying astrophysical parameters, $f_{*,10}$, $f_\mr{esc, 10}$, and $M_\mr{turn}$, from top to bottom. The values for these parameters assumed here (shown inside the figure) correspond to the mean, 1$\sigma$ upper and lower values from the HERA results \cite{2022ApJ...924...51A}. }
    \label{fig:IonHist_3astros}
\end{figure}

When calculating the reionization history, we use the astrophysical parameters adopted in the \texttt{21cmFAST} version~2 \cite{2019MNRAS.484..933P}\footnote{The latest version of \texttt{21cmFAST} is 3.2.1 (updated September 13, 2023), which has improved calculation speed and numerical stability. However, version 2 is slightly easier in extension and does not affect the calculation results. Therefore we use version 2 in this paper.}. In particular, three parameters $f_{*,10}$, $f_\mr{esc, 10}$, and $M_\mr{turn}$ give a significant impact on the reionizaiton history.
These model parameters are defined by the stellar mass-to-baryon mass fraction in halos
\begin{align}
    f_{*}(M_\mr{halo}) = f_{*,10} \left(\cfrac{M_\mr{halo}}{10^{10} M_\odot}\right)^{\alpha_*},
\end{align}
the halo mass-dependent escape fraction of the ionizing photons
\begin{align}
    f_\mr{esc}(M_\mr{halo}) = f_\mr{esc,10} \left(\cfrac{M_\mr{halo}}{10^{10} M_\odot}\right)^{\alpha_\mr{esc}},
\end{align}
and the duty cycle of the star-forming galaxy
\begin{align}
    f_\mr{duty}(M_\mr{halo}) = \exp \left(-\cfrac{M_\mr{turn}}{M_\mr{halo}}\right) \,,
\end{align}
where $\alpha_*$ and $\alpha_\mr{esc}$ represent their power-law dependence.
Figure \ref{fig:IonHist_3astros} shows the time evolution of the free electron fraction with different astrophysical parameters, $f_{*,10}$, $f_\mr{esc, 10}$, and $M_\mr{turn}$. Other astrophysical parameters are set to the best-fitted values of HERA results \cite{2022ApJ...924...51A} as $\alpha_{*}=0.50$, $\alpha_\mr{esc}=0.02$, $t_\mr{*}=0.60$, $L_\mr{X}/\mr{SFR}=10^{40.64}~\mr{erg/s/M_\odot~yr}$.
These parameters can also affect the reionization history. The effects on the free electron fraction at $z=7.3$ are $\Delta x_\mr{e} \lesssim 0.01$ for varying $t_\mr{*}$, $\Delta x_\mr{e} \lesssim 0.05$ for varying $L_\mr{X}/\mr{SFR}$. Thus, these effects are smaller than the above three, $f_{*,10}$, $f_\mr{esc, 10}$, and $M_\mr{turn}$. On the other hand, $\Delta x_\mr{e}$ can be larger than 0.1 for varying $\alpha_{*}$ and $\alpha_\mr{esc}$. However, we only use the free electron fraction at $z=7.3$ and it is difficult to constrain the $\alpha_{*}$ and $\alpha_\mr{esc}$ independently of $f_{*,10}$ and $f_\mr{esc, 10}$, respectively. Thus, we also fix $\alpha_{*}$ and $\alpha_\mr{esc}$.
For the simulation parameters in \texttt{21cmFAST}, we take such that the box size is (400 Mpc)$^3$, spatial resolution is $800^3$ for high-resolution calculation, and $200^3$ for low-resolution.

Our extended \texttt{21cmFAST} enables us to calculate the reionization history varying the primordial running spectral indices and astrophysical parameters. However, the parameter inference with MCMC analysis, which is widely used to obtain constraints in cosmology and reionization \cite{2015MNRAS.449.4246G,2020PhRvD.101f3526M,2022ApJ...924...51A,2022MNRAS.516.5588G}, requires computational resources so much\footnote{In Ref. \cite{2022arXiv221214064N}, using 21CMMC\cite{2015MNRAS.449.4246G}, they performed an MCMC analysis with varying two astrophysical parameters and two parameters related to the bump-like features of the primordial power spectrum. They have used the Planck optical depth constraint to perform the MCMC analysis, but have not discussed other observations of reionization.}. To save the calculation time, we construct the fitting function for the free electron fraction at redshift $z=7.3$, based on our extended \texttt{21cmFAST} results.
Here, we use the free electron fraction only at $z=7.3$, at which the most stringent constraint is given by the SILVERRUSH data.
The fitting function that we obtained is
\begin{align}
x_\mr{e}(z=7.3)=\min (1.0, \tilde{x}_\mr{e, 7.3}),
\label{eq:fitting_function1}
\end{align}
where
\begin{align}
\tilde{x}_\mr{e, 7.3} = &\left\{0.421 \mr{e}^{A}
+B\left[\left(\cfrac{f_{*,10}}{0.058}\right)^{C}-1.0\right] 
+D \log \left(\cfrac{M_{\mr{turn}}}{3.8 \times 10^8 M_{\odot}}\right)\right\}
\left(\cfrac{f_{\mr{esc,10}}}{0.078}\right)^{E},
\label{eq:fitting_function2}
\end{align}
with
\begin{align}
A&=8.43 \alpha_s+11.41 \beta_s,\nonumber \\
B&=4.00 \alpha_s+7.75 \beta_s+0.38,\nonumber \\
C&=-5.24 \alpha_s-12.04 \beta_s+1.34,\nonumber \\
D&=-0.7 \alpha_s-1.20 \beta_s-0.07,\nonumber \\
E&=-3.11 \beta_s+1.08.\nonumber 
\end{align}

In order to check the validity of our fitting function, we performed test calculations with the random parameter sets ($\alpha_s, \beta_s, f_{\mr{esc,10}}, f_{*,10}, M_{\mr{turn}}$) in the ranges of $\alpha_s = (-0.05,0.05), \beta_s=(-0.05,0.05), f_{\mr{esc,10}}=(0.03,0.3), f_{*,10}=(0.024,0.091),$ and $M_{\mr{turn}}/ M_\odot=(1.6\times 10^8,1.5\times 10^9)$.
We show the comparison between the extended \texttt{21cmFAST} results and the fitting function in Table \ref{tab:fitting_function}. 
The absolute value of the average difference of the free electron fraction between that from \texttt{21cmFAST} and the fitting function $\Delta x_\mr{e}$ is about 0.02, which is well smaller than the error in the SILVERRUSH data. Therefore we can safely use our fitting function to analyze the constraint.
Our fitting function \eqref{eq:fitting_function2} is obtained from the simulation data from \texttt{21cmFAST} with $\alpha_s=(-0.015,0.02), \beta_s=(-0.015,0.02), f_{\mr{esc,10}}=(0.03,0.3), f_{*,10}=(0.024,0.091), M_{\mr{turn}}/M_\odot=(1.6\times 10^8,1.5\times 10^9)$. However, as seen in the results of Table~\ref{tab:fitting_function}, the fitting function can also be applied outside the range of parameters that we used. 
Although one may expect that we can also use other parameters such as $n_s, \alpha_*, \alpha_\mr{esc}$ to improve the accuracy of our fitting function, it turns out that the inclusion of more parameters in the fitting does not help much. Therefore, we decided to fix other parameters to obtain the fitting function.

\begin{table}
	\centering
	\caption{Comparison between the free electron fraction calculated with the extended \texttt{21cmFAST} code and our fitting function Eq.~\eqref{eq:fitting_function2}. The average value of the absolute difference between the \texttt{21cmFAST} and Eq.~\eqref{eq:fitting_function2} is about 0.02 among the parameter sets we used.}
        \vspace{4mm}
	\label{tab:fitting_function}
 	\begin{tabular}{|rrrrr|ccr|} % four columns, alignment for each
		\hline
		$\alpha_s$ & $\beta_s$ & $f_\mr{esc,10}$ & $f_{*,10}$ & $M_\mr{turn}/10^8 M_\odot$ & $x_\mr{e}^\mr{sim}(z=7.3)$ & $x_\mr{e}^\mr{fit}(z=7.3)$ & $\Delta x_\mr{e}$\\
		\hline
        0.002 & 0.031 & 0.074 & 0.055 & 10.240 & 0.442 & 0.451 & -0.009 \\
        -0.038 & 0.015 & 0.113 & 0.067 & 3.857 & 0.688 & 0.650 & 0.038 \\
        -0.005 & -0.013 & 0.124 & 0.067 & 3.788 & 0.748 & 0.705 & 0.042 \\
        -0.034 & -0.000 & 0.064 & 0.087 & 6.155 & 0.383 & 0.409 & -0.025 \\
        0.017 & -0.030 & 0.100 & 0.079 & 4.691 & 0.688 & 0.645 & 0.043 \\
        0.032 & 0.039 & 0.050 & 0.068 & 2.139 & 0.684 & 0.676 & 0.008 \\
        0.007 & -0.002 & 0.086 & 0.074 & 2.164 & 0.718 & 0.704 & 0.014 \\
        0.005 & 0.019 & 0.166 & 0.033 & 2.845 & 0.730 & 0.724 & 0.007 \\
        0.011 & 0.011 & 0.149 & 0.042 & 5.242 & 0.729 & 0.666 & 0.063 \\
        0.009 & -0.018 & 0.100 & 0.055 & 5.932 & 0.447 & 0.437 & 0.011 \\
		\hline
	\end{tabular}
\end{table}

\subsection{Constraints from the SILVERRUSH observation}
\label{sec:2-3}
In this study, we perform MCMC analysis, using the public code {\tt emcee} \cite{2013PASP..125..306F}, by comparing the SILVERRUSH results with Eq.~\eqref{eq:fitting_function2} using a Gaussian likelihood. 
We use two priors on the running parameters: the flat one with $\alpha_s=(-0.1,0.1), \beta_s=(-0.1,0.1)$, and the Gaussian Planck prior.
The Planck prior assumes the 2-dimensional Gaussian distribution function with the mean $\alpha_s=0.0012, \beta_s=0.0119$ and the covariance matrix from the Planck Legacy Archive \cite{PlanckLegacyArchive}. Three astrophysical parameters are varied with flat priors as $f_{\mr{esc,10}}=(0.03,0.3), f_{*,10}=(0.024,0.091),$ and $\log_{10}(M_\mr{turn}/M_\odot)=(8.2,9.2)$, which correspond to the 1$\sigma$ uncertainty range from the HERA result \cite{2022ApJ...924...51A}.

\begin{figure*}
    \centering
	\includegraphics[width=0.7\columnwidth]{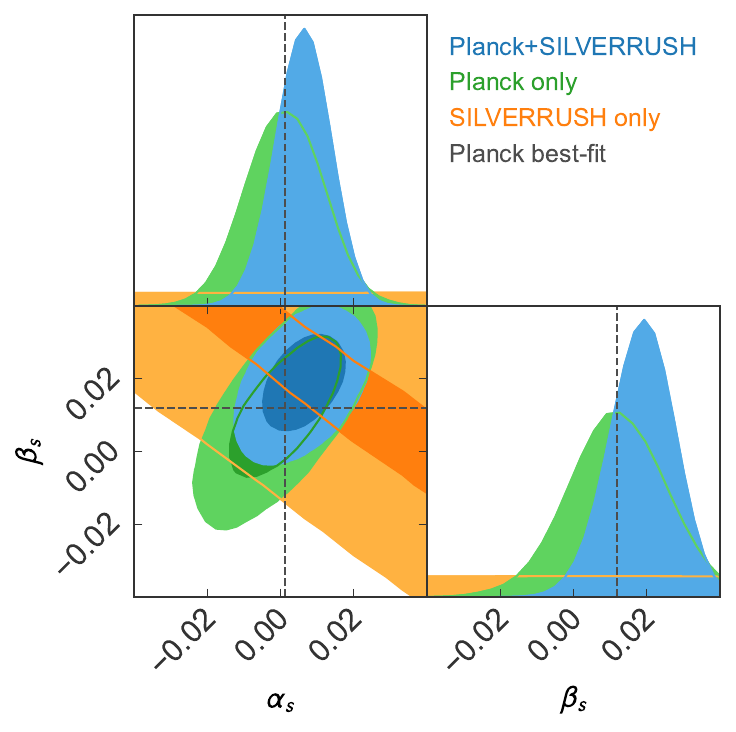}
    \caption{Triangle-plot of constraints on $(\alpha_s, \beta_s)$ from observations of the reionization history. The constraint with the flat prior on $(\alpha_s, \beta_s)$ is labeled as ``SILVERRUSH only'' (orange), and one with the Gaussian Planck prior is labeled as ``Planck+SILVERRUSH'' (blue). For comparison, the constraint obtained from Planck results is also shown as ``Planck only'' (green) \cite{2020A&A...641A..10P, PlanckLegacyArchive}.
    }
    \label{fig:const2D_onlyRunning_SilverRush}
\end{figure*}

First, we show constraints on $\alpha_s$ and $\beta_s$ for the case that the astrophysical parameters are fixed as $f_\mr{esc,10}=0.078, f_{*,10}=0.058,$ and $\log_{10}(M_\mr{turn}/M_\odot)=8.58$, which are taken from the recent HERA best-fitted values in Figure \ref{fig:const2D_onlyRunning_SilverRush}. When imposing the flat priors for $\alpha_s$ and $\beta_s$, a strong degeneracy between $\alpha_s$ and $\beta_s$ appears. This degeneracy shows a negative correlation because the reionization redshift is considered to be determined by the small-scale matter power spectrum where effects of $\alpha_s$ and $\beta_s$ are canceled.
On the other hand, CMB anisotropies probe large-scale perturbations, and therefore the direction of the degeneracy is positive since the response of large-scale fluctuations to the change of $\alpha_s$ and $\beta_s$ is opposite to that of small-scale as seen from Figure~\ref{fig:PSmat_running_ab}.
Figure \ref{fig:const2D_onlyRunning_SilverRush} shows that combining CMB and reionization observations can break these degeneracies and bring stringent constraints on $\alpha_s$ and $\beta_s$.

\begin{figure*}
    \centering
	\includegraphics[width=0.7\columnwidth]{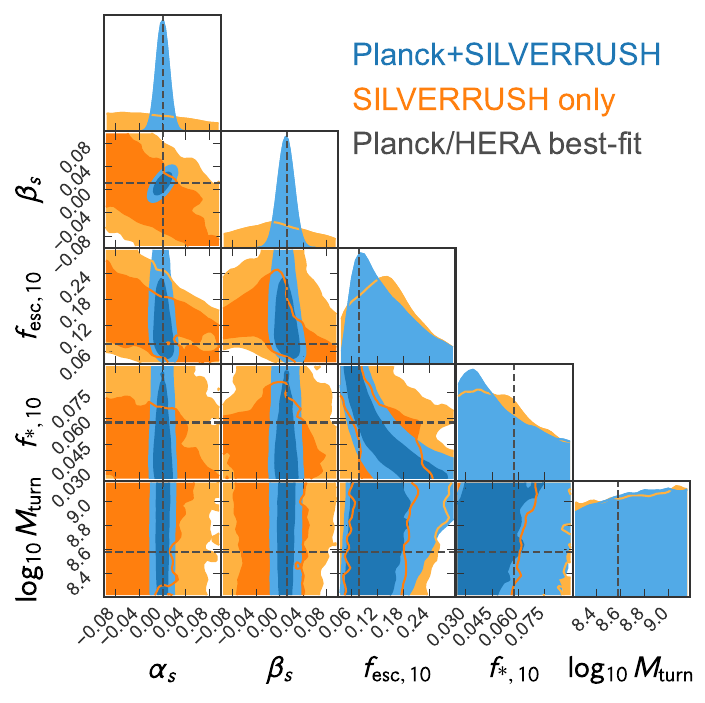}
    \caption{Triangle-plot of constraint on $(\alpha_s, \beta_s, f_\mr{*,10}, f_\mr{esc,10}, M_\mr{turn})$ from the SILVERRUSH observations with the flat prior on $(\alpha_\mr{s},\beta_\mr{s})$ (orange), and that with the Gaussian Planck prior \cite{PlanckLegacyArchive} (blue).
    }
    \label{fig:const2D_all_silverrush_planck}
\end{figure*}

When one discusses the constraints on the running spectral indices from the reionization history, the degeneracies between the running and astrophysical parameters should be considered. Figure \ref{fig:const2D_all_silverrush_planck} shows the results of MCMC analysis when all parameters are sampled. 
When we assume a flat prior on $\alpha_s$ and $\beta_s$, although one can see a weak correlation between $\alpha_s$ and $\beta_s$, most parameters are poorly constrained and the 68\% C.L. errors on the running spectral indices are $\Delta \alpha_s \sim \Delta \beta_s \sim {\cal O}(0.1)$.
On the other hand, when we impose the Planck prior for $\alpha_s$ and $\beta_s$, the constraints become significantly improved as $\Delta \alpha_s \sim \Delta \beta_s \sim 0.01$ while constraints on astrophysical parameters are not changed much between those for the two different priors. 
Furthermore, we found that $f_\mr{esc,10}$ and $f_{*,10}$ are slightly more sensitive to the reionization history, compared with $M_\mr{turn}$ as expected from Figure \ref{fig:IonHist_3astros}. In table \ref{tab:EoR_constraints}, we summarize the best-fit values and 68\% C.L. errors for two prior cases with and without fixing astrophysical parameters.

\begin{table}
	\centering
	\caption{The best-fit values and 68\% C.L. errors for the running and astrophysical parameters from SILVERRUSH observations.
	\label{tab:EoR_constraints}} \vspace{4mm}
	\begin{tabular}{|c|c|c|c|c|} % four columns, alignment for each
	    \hline
	    & \multicolumn{2}{|c|}{w/o astrophysical parameters} & \multicolumn{2}{c|}{w/ astrophysical parameters}\\
		\hline
		& flat prior & Planck prior & flat prior & Planck prior \\
		\hline
        $\alpha_s$ & --- & $0.006^{+0.007}_{-0.007}$ & $-0.048_{-0.103}^{+0.149}$ & $0.001_{-0.010}^{+0.010}$\\
        $\beta_s$ & --- & $0.019^{+0.008}_{-0.009}$ & $0.048_{-0.118}^{+0.110}$ & $0.012_{-0.013}^{+0.013}$\\
        $f_\mr{esc,10}$ && & $0.087_{-0.064}^{+0.147}$ & $0.112_{-0.041}^{+0.055}$\\
        $f_{*,10}$ && & $0.104_{-0.072}^{+0.182}$ & $0.064_{-0.024}^{+0.022}$\\
        $\log_{10}(M_\mr{turn}/M_\odot)$ && & $9.148_{-1.370}^{+0.937}$ & $9.053_{-0.737}^{+0.912}$\\
		\hline
	\end{tabular}
\end{table}

\subsection{Effects of the redshift width of reionization $\Delta z$ \label{sec:2-4}}
% \subsection{Effects of the redshift width of reionization}

\begin{figure*}
    \centering
	\includegraphics[width=0.7\columnwidth]{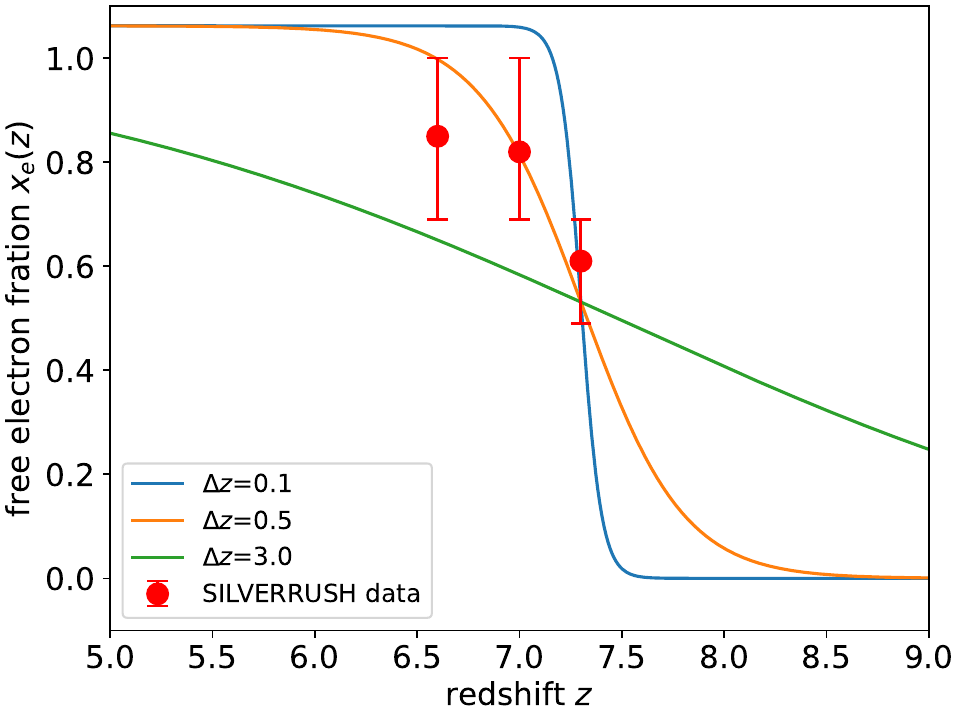}
    \caption{Reionization history with different values of $\Delta z=0.1, 0.5,$ and 3.0. For all cases, the free electron fraction at $z=7.3$ is fixed at $x_\mr{e}(z=7.3)=0.5$.
    }
    \label{fig:different_deltaz}
\end{figure*}

So far, we have fixed the redshift width of $\tanh$-type reionization as $\Delta z=0.5$.
Here, we compare the MCMC analyses with fixed $\Delta z$ and with varied $\Delta z$.
In the analysis so far, we used only one redshift data point for SILVERRUSH, as $0.49 \le x_\mr{e} \le 0.69$ at $z=7.3$.
When fixing all the parameters, the free electron fraction at $z=7.3$ is determined by using Eqs. \eqref{eq:fitting_function1} and \eqref{eq:fitting_function2}. 
Since the analysis with only one redshift data cannot constrain the effect of varying $\Delta z$, we use three different redshift datapoints of SILVERRUSH survey, as $x_\mr{e} \ge 0.69$ at $z=6.6$, $x_\mr{e} \ge 0.69$ at $z=7.0$, and $0.49 \le x_\mr{e} \le 0.69$ at $z=7.3$ in this section. In the MCMC analysis with varied $\Delta z$, after determining the $x_\mr{e}$ at $z=7.3$ by choosing model parameters $(\alpha_s, \beta_s, f_\mr{*,10}, f_\mr{esc,10}, M_\mr{turn})$, the $x_e$ at the other redshifts (i.e., at $z=6.6$ and $z=7.0$ can be determined by giving $x_\mr{e} (z=7.3)$ and $\Delta z$ based on the $\tanh$-type function \eqref{eq:tanh_reionization}. 
For reference, we show the reionization history with different $\Delta z$ in Figure \ref{fig:different_deltaz}. This figure shows that the additional two data points may be useful to constrain the width of reionization $\Delta z$.

\begin{figure*}
    \centering
	\includegraphics[width=0.7\columnwidth]{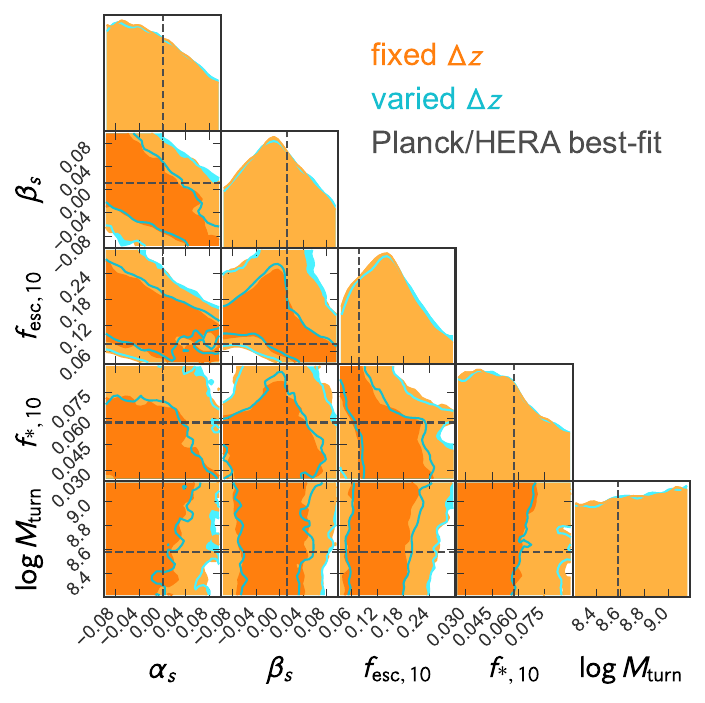}
    \caption{Triangle-plot of constraint on $(\alpha_s, \beta_s, f_\mr{*,10}, f_\mr{esc,10}, M_\mr{turn})$ from the SILVERRUSH observations with $\Delta z$ being fixed (blue) and varied (orange). Here we assume a flat prior for all parameters.
    }
    \label{fig:const2D_all_silverrush_varyingdz}
\end{figure*}
\begin{figure*}
    \centering
	\includegraphics[width=0.7\columnwidth]{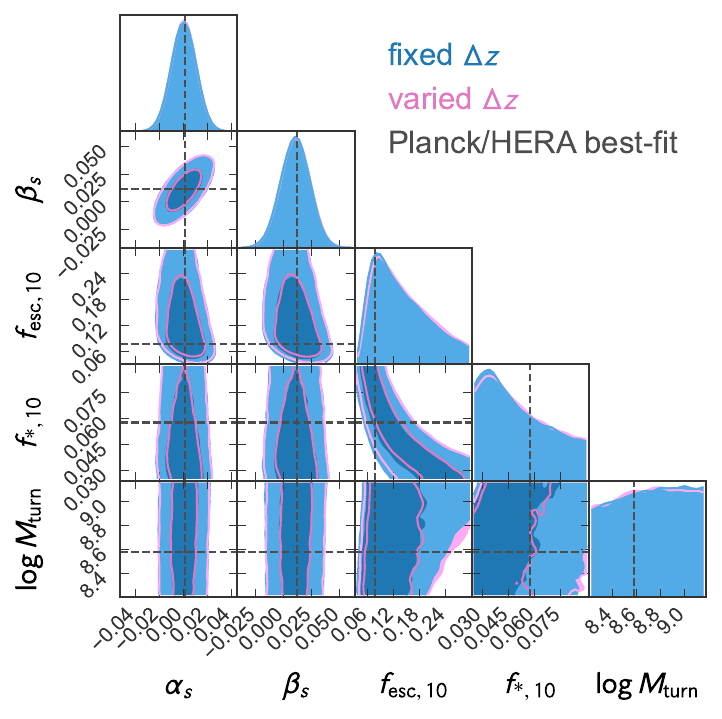}
    \caption{Triangle-plot of constraint on $(\alpha_s, \beta_s, f_\mr{*,10}, f_\mr{esc,10}, M_\mr{turn})$ from the SILVERRUSH observations with $\Delta z$ being fixed (blue) and varied (orange). Here we assume the Gaussian Planck prior for $\alpha_s$ and $\beta_s$, but a flat prior on other parameters.
    }
    \label{fig:const2D_all_silverrush_planck_varyingdz}
\end{figure*}

In the analysis with varied $\Delta z$, we impose the flat prior distribution as $\Delta z=(0.1,3.0)$. This prior choice is motivated by the following reasons. The observation of the kinetic Sunyaev-Zel'dovich effect indicates $\Delta z<0.79$ at 95\% C.L. \cite{2016A&A...596A.108P}. On the other hand, the radiation-hydrodynamic (RHD) simulation predicts relatively higher values of $\Delta z \sim 1.5-2.5$ \cite{2018ApJ...858L..11T}. It should be noted that, however, since the RHD simulation results do not follow the $\tanh$-type redshift evolution, it is not straightforward to convert their results to the constraints on $\Delta z$. Therefore we adopt a conservative prior on the $\Delta z$.

Figure \ref{fig:const2D_all_silverrush_varyingdz} shows the comparison of constraints from SILVERRUSH observations with flat priors on all parameters $(\alpha_s, \beta_s, f_\mr{*,10}, f_\mr{esc,10}, M_\mr{turn})$, with $\Delta z$ being fixed and varied.
We found that $\Delta z$ is constrained as $\Delta z = 1.327_{-0.754}^{+1.056}$ at 68\% C.L.  Constraints on the other parameters in the analysis with and without varied $\Delta z$ are almost the same. 
In Figure \ref{fig:const2D_all_silverrush_planck_varyingdz}, we show the case with the Gaussian Planck prior on $(\alpha_\mr{s},\beta_\mr{s})$, but with a flat prior for the others $(f_\mr{*,10}, f_\mr{esc,10}, M_\mr{turn})$.
In this analysis, we obtained $\Delta z = 1.339_{-0.767}^{+1.043}$ at 68\% C.L. and constraints on other parameters are almost unchanged even when we vary $\Delta z$ as in Figure~\ref{fig:const2D_all_silverrush_varyingdz}.
From the results of Figures~\ref{fig:const2D_all_silverrush_varyingdz} and \ref{fig:const2D_all_silverrush_planck_varyingdz}, we can conclude that varying redshift width of reionization $\Delta z$ does not give a significant impact on the constraints on the primordial fluctuations in our analysis.

\section{Conclusion}
\label{sec:3}
In this work, we have argued that the reionization history can be used as a novel probe of primordial fluctuations. To this end, we have investigated the impact of the small-scale primordial curvature perturbations on the reionization history. We modified the semi-numerical simulation code \texttt{21cmFAST} to examine the reionization history incorporating the running of the spectral indices of the primordial power spectrum up to the second order.
Based on calculations with \texttt{21cmFAST}, we constructed the fitting function for the free electron fraction at $z=7.3$. We also performed the MCMC analysis by using the fitting function and observational data of the reionization history from the SILVERRUSH LAEs to obtain the constraints on the running spectral indices. In case the astrophysical uncertainties are neglected, the running spectral indices can be strongly constrained by combining the observation of the reionization history and that of CMB. 
On the other hand, when considering uncertainties of the astrophysical parameters as inferred in the recent 21-cm observations \cite{2022ApJ...924...51A}, we found that the current observational constraints on the reionization history cannot severely constrain the running spectral indices.
This result is based on the relatively simple astrophysical model, which is adopted in the \texttt{21cmFAST}. Therefore, if one considers a more complicated astrophysical model, the constraints on the running parameters would get looser.
However, in near future, the measurements of astrophysical parameters can become more precise and, in such a case, the reionization history would be a useful probe of primordial fluctuations and can augment other methods to severely constrain them.

Finally, we comment on the other observations of the reionization history. In this work, we mainly used the free electron fraction derived from the luminosity function of the LAEs observed in the SILVERRUSH survey. Our choice of observational data is just to demonstrate the power of constraining primordial perturbations using the reionization history.
Actually, many observational constraints on the reionization history are reported, such as the Ly-$\alpha$ equivalent width of the Lyman break galaxies (LBGs) \cite{2018ApJ...856....2M,2019MNRAS.485.3947M,2019ApJ...878...12H,2020MNRAS.495.3602W}, the number fraction of LBGs emitting Ly-$\alpha$ in all observed LBGs \cite{2015MNRAS.446..566M}, gamma-ray burst and the QSO damping wings \cite{2006PASJ...58..485T,2014PASJ...66...63T,2013MNRAS.428.3058S,2018ApJ...864..142D,2019MNRAS.484.5094G,2020ApJ...896...23W}, the Ly-$\alpha$ and Ly-$\beta$ forest dark gaps of QSO spectra \cite{2015MNRAS.447..499M,2022ApJ...932...76Z}, and the Gunn-Peterson trough of QSOs \cite{2006AJ....132..117F,2011MNRAS.415L...1G}.
These observational data are independent but some of the above results show inconsistencies (for examples, see Figure 11 of Ref. \cite{2021ApJ...923..229G}). Among them, the reionization history obtained by the LAE luminosity function with the Subaru telescope that we used in this work provides constraints at different redshifts, and they are consistent with each other \cite{2014ApJ...797...16K,2018PASJ...70...55I,2018ApJ...867...46I,2018PASJ...70S..16K,2021ApJ...923..229G}.
It is worth noting that measurements of the LAE luminosity function using other telescopes provide different constraints on the reionization history \cite{2019ApJ...886...90H,2021ApJ...919..120M}, and therefore the best-fit values in our analysis may depend on the observational data choice.

The current observational constraint errors on the free electron fractions are $\Delta x_\mr{e} \sim \mathcal{O}(0.1)$. However, we would expect that more precise measurements of the reionization history can decrease the uncertainties on cosmological and astrophysical parameters.
However, it may be difficult to resolve degeneracies among the parameters in our analysis, i.e., $\alpha_s, \beta_s, f_\mr{*,10}, f_\mr{esc,10},$ and $M_\mr{turn}$ by using observations of the reionization history only. Further improvement on the constraints on the primordial curvature perturbations is expected to be obtained by combining other types of observations, such as the recent high-redshift galaxy luminosity function obtained by the James Webb Space Telescope \cite{2022arXiv220712446L,2022arXiv220814999E,2023MNRAS.518.4755A}, which is worth investigating. We leave this issue to future work.

%\clearpage
\section*{Acknowledgements}
T. M. is supported by JSPS Overseas Research Fellowship and Shui Mu Fellowship. S. Y. is supported by JSPS Research Fellowships for Young Scientists. This work was supported in part by JSPS KAKENHI Grant Nos. 21J00416~(S. Y.), 22KJ3092~(S. Y.) and 19K03874~(T. T.). Numerical computations were carried out on Cray XC50 at Center for Computational Astrophysics, National Astronomical Observatory of Japan.

\clearpage 
\bibliography{EOR_running}

\end{document}